\documentclass[10pt,showpacs,twocolumn,preprintnumbers,amsmath,amssymb,
aps,prd,nofootinbib,eqsecnum,a4paper]{revtex4}

\usepackage{epsfig}
\usepackage{graphicx,epsf}
\usepackage{color}
\usepackage{bm}
\usepackage{psfrag}
\def\be{\begin{equation}}
\def\ee{\end{equation}}
\def\beb{\begin{equation*}}
\def\eeb{\end{equation*}}
\def\bea{\begin{eqnarray}}
\def\eea{\end{eqnarray}}
\def\beab{\begin{eqnarray*}}
\def\eeab{\end{eqnarray*}}
\def\nn{\nonumber}

\def\Z{{{\cal{Z}}}}

\def\vp{{{\varphi}}}
\def\dvp{{\delta\varphi}}

\def\1{{}^{(1)}}
\def\2{{}^{(2)}}
\def\cs2{c_{\rm{s}}^2}

\def \beg {\begin{enumerate}}
\def \en {\end{enumerate}}

\def\vpb{\varphi_0}

\def\cs{c_{\rm{s}}^2}

\def\dm{{\delta_{\rm c}}}

\begin{document}
%%%%%%%%%%%%%%%%%%%%%%

\title{Gauge conditions in combined dark energy and dark matter systems}
\author{Adam J.~Christopherson}
\email{a.christopherson@qmul.ac.uk}
\affiliation{
Astronomy Unit, School of Mathematical Sciences, Queen Mary University
of London, Mile End Road, London, E1 4NS, United Kingdom}
\date{\today}

%\[\]
%\end{center}
\begin{abstract}
%, there has been much interest in systems consisting of both dark matter
%and dark energy. 
When analysing a system consisting of both dark matter and dark energy,
an often used practice 
in the literature is
to neglect the perturbations in the dark energy component.
However, it has recently been argued, through
the use of numerical simulations, that one cannot do so. In this work
we show that by neglecting such perturbations one is implicitly making a choice of gauge.
 As such, one no longer has the freedom to choose, for example,
a gauge comoving with the dark matter -- in fact doing so will give 
erroneous, gauge dependent results. We obtain results consistent with the numerical
simulations by using the formalism of cosmological
perturbation theory, and thus without resorting to involved numerical calculations.
\end{abstract}


\maketitle
%%%%%%%%%%%%%%%%%%%%%%%%%%%%%%%%%%%%%
\section{Introduction}

Cosmological data sets have increased in both their number and quality
in recent years and we now hold a wealth of data about our universe.
 This allows us 
to confront the theory with observations and thus obtain
a knowledge of the intricacies of the universe and its evolution 
better than ever before. Current data allows us to conclude that
the majority of the energy density of the universe today is in the 
so called `dark sector'. That is, the universe is currently
predominantly made up of  dark energy -- the mysterious 
force responsible for the accelerated expansion of the universe
at the current epoch -- and dark matter 
\cite{WMAP7, Seljak:2006bg, Tegmark:2006az, Reid:2009xm}. 

The nature of dark energy is arguably one of the most important open
questions in modern cosmology. An obvious `candidate' for
the dark energy component is the cosmological constant, $\Lambda$. In fact, current 
observations are consistent with $\Lambda$CDM, the standard
cosmological model in which the dark energy is a cosmological 
constant \cite{WMAP7}. However, in light of the `cosmological constant problem,'
that the observed value of $\Lambda$ differs from that predicted 
from fundamental theories by over a hundred orders of magnitude 
\cite{Copeland:2006wr},
the cosmological constant is perhaps a less favourable candidate for dark energy.
Thus, much recent work has been
focussed on a dynamical agent for the dark energy with a 
negative pressure which will cause the universe to accelerate in
its expansion (see Ref.~\cite{K} and references contained within).\\

One of the most useful tools that theoretical cosmologists
have at their disposal is cosmological perturbation theory.
The basic premise is simple: since the universe appears homogeneous
and isotropic on large scales, one starts with a 
Friedmann-Robertson-Walker spacetime as a background solution, and 
then adds small inhomogeneous perturbations on top. (For 
a selection of papers and reviews on the topic see, e.g., Ref.~\cite{Kodama:1985bj, Mukhanov:1990me, 
Bertschinger:1993xt, 
Riotto:2002yw, durrerbook}). Since one perturbs the geometry of the
background spacetime, through Einstein's field equations, this invokes
perturbations on its matter content. For a universe whose
dark energy component is pure cosmological constant only the dark matter
will be perturbed, but in the case of dynamical dark energy there will be
perturbations to both the dark matter and the dark energy components. However, an often used 
practice in the literature (e.g. Ref.~\cite{Huterer:2006mva} and see references in \cite{Hwang:2009zj}) is to
ignore the perturbations in the dark energy and simply use the well-known
evolution equation given by 
\be 
\label{eq:standard}
\ddot{\dm}+2H\dot{\dm}-4\pi G \rho_{\rm c}\dm
=0\,,
\ee
where $\dm\equiv\delta\rho_{\rm c}/\rho_{\rm c}$ is the dark matter density contrast
and an overdot denotes a time derivative. This issue was addressed in Ref.~\cite{Hwang:2009zj},
where Park et. al. showed, using numerical methods, that dark energy perturbations
are not negligible and so, in general, cannot be ignored. In this paper we address a similar
question. Using the formalism of cosmological perturbation theory, we show that  
ignoring the dark energy perturbations is in fact a gauge choice, and so if 
we choose to set the dark energy perturbations to zero then we no longer have
the freedom to choose a gauge in which the dark matter evolution equation takes the
form of Eq.~(\ref{eq:standard}) -- we are still left with a system of coupled differential
equations. Since this work focusses on the formalism of cosmological perturbation
theory, and mathematics, and not involved numerical calculations,
we make no numerical approximations in obtaining this result.
\\

The paper is structured as follows. In the next section we derive and present the
governing equations for the dark energy and dark matter system
 under consideration. This is followed, in
Section \ref{sec:gaugetrans}, by a brief recap of the gauge transformations of 
matter and metric perturbations.
Finally, in Section \ref{eq:conc} we present our results and conclude.

\section{governing equations}

We consider here only scalar perturbations to a flat ($K=0$) Friedmann-Robertson-Walker
spacetime with the line element
\bea
ds^2&=&-(1+2\phi)dt^2+2a^2B_{,i}dt dx^i\nn\\
&&+a^2[(1-2\psi)\delta_{ij}+2E_{,ij}]dx^idx^j\,,
\eea
where $a=a(t)$ is the scale factor, $\phi$ is the lapse function, $\psi$ is 
the dimensionless curvature perturbation and $E$ and $B$ make up the 
scalar shear, $\sigma$, as
\be 
\sigma\equiv a^2\dot{E}-aB\,.
\ee
All perturbations are functions of space and time (e.g. $\phi\equiv\phi(x^i,t)$).
Throughout this paper, Latin indices $i,j,k,$ take the value $1$, $2$ or $3$,
Greek indices $\mu,\nu,\lambda,$ cover the full spacetime indices and we 
denote a partial derivative
with a subscript comma.

We assume that the dark matter and dark energy are non-interacting, and 
so energy momentum conservation for each fluid gives\footnote{
This assumption is for brevity, and to allow us to deal with more
manageable equations. The results highlighted in this work will still
hold in the case of interacting fluids for which
the overall energy-momentum tensor is covariantly conserved, but the components
obey
\be 
\nabla_\mu T_{(\alpha)}^\mu{}_\nu=Q_{(\alpha)\nu}\,,
\ee
where $Q_{(\alpha)\nu}$ is the energy-momentum transfer to the $\alpha$th fluid \cite{Hwang:2001fb, 
MW2008}.
}
\be 
\label{eq:emcons}
\nabla_\mu T_{(\alpha)}^\mu{}_\nu=0\,.
\ee
Then, we obtain an evolution equation for each fluid from the energy (temporal) component of 
Eq.~(\ref{eq:emcons}),
\begin{align}
\label{eq:evgen}
\dot{\delta\rho}_\alpha&+3H(\delta\rho_\alpha+\delta P_\alpha)\nn\\
&
+(\rho_\alpha+P_\alpha)\frac{\nabla^2}{a^2}(V_\alpha+\sigma)
%\qquad\qquad\qquad\qquad\qquad\qquad
=3(\rho_\alpha+P_\alpha)\dot{\psi}\,,
\end{align}
where the covariant velocity potential of each fluid is defined as
\be 
\label{eq:covvel}
V_\alpha=a(v_\alpha+B)\,,
\ee 
$v_\alpha$ is the scalar velocity potential of the $\alpha$th fluid and 
$H=\dot{a}/a$ is the Hubble parameter.

Considering the dark matter fluid and the dark energy scalar field, respectively, 
Eq.~(\ref{eq:evgen}) then gives
\bea
\label{eq:fluidev}
&&\dot{\dm}=3\dot{\psi}
-\frac{\nabla^2}{a^2}(\sigma+V_{\rm c})\,,\\
\label{eq:fieldev}
&&\ddot{\dvp}+3H\dot{\dvp}+\Big(U_{,\vp\vp}-\frac{\nabla^2}{a^2}\Big)\dvp\nn\\
&&\qquad\qquad=\dot{\vp}\Big(3\dot{\psi}-\frac{\nabla^2}{a^2}\sigma+\dot{\phi}\Big)-2U_{,\vp}\phi\,,
\eea
where we have treated the field as a perfect fluid with energy density and pressure
\bea
\rho_\vp=\frac{1}{2}\dot{\vp}^2+U_{,\vp}\,,\\
P_\vp=\frac{1}{2}\dot{\vp}^2-U_{,\vp}\,.
\eea
We have also
used the background evolution equation for the dark energy scalar field
\be 
\ddot{\vp}+3H\dot{\vp}+U_{,\vp}=0\,,
\ee
and the fact that
\be 
V_\vp=-\frac{\dvp}{\dot{\vpb}}\,.
\ee 
There is also a momentum conservation equation, coming from the 
spatial component of Eq.~(\ref{eq:emcons}), corresponding to each fluid
\be 
\label{eq:mmtmcons}
\dot{V_\alpha}-3H c_\alpha^2 V_\alpha+\phi+\frac{\delta P_\alpha}{\rho_\alpha +P_\alpha}=0\,,
\ee
where $c_\alpha^2=\dot{P_\alpha}/\dot{\rho_\alpha}$.
\\
The Einstein field equations give \cite{Malik:2004tf}
\begin{align}
\label{eq:00}
&3H(\dot{\psi}+H\phi)-\frac{\nabla^2}{a^2}(\psi+ H \sigma)+4\pi G \delta\rho=0\,,\\
\label{eq:0i}
&\dot{\psi}+H\phi+4\pi G(\rho+P)V=0\,,\\
\label{eq:ijtracefree}
&\dot{\sigma}+H\sigma-\phi+\psi=0\,,\\
\label{eq:ijtrace}
&\ddot{\psi}+3H\dot{\psi}+H\dot{\phi}+(3H^2+2\dot{H})\phi
-4\pi G\delta P=0\,,
\end{align}
where the total matter quantities are defined as the sum of the quantity for each fluid/field, i.e.
\begin{align}
\delta\rho&=\delta\rho_\vp+\delta\rho_{\rm c}\,,\\
\delta P&=\delta P_{\vp}\,,\\
(\rho+P)V&=(\rho_\vp+P_\vp)V_\vp+\rho_{\rm c}V_{\rm c}\,,
\end{align}
and we have used the fact that the dark matter is pressureless, i.e.
$P_{\rm c}=0=\delta P_{\rm c}$.
Note that, by treating the scalar field as a fluid, we can write 
the energy density and pressure perturbation for the dark energy, respectively,
as
\begin{align}
\delta\rho_\vp&=\dot{\vp}\dot{\dvp}-\phi\dot{\vp}^2+U_{,\vp}\dvp\,,\\
\delta P_{\vp}&=\dot{\vp}\dot{\dvp}-\phi\dot{\vp}^2-U_{,\vp}\dvp\,.
\end{align}
Introducing a new variable $\Z$, both for notational convenience and
to assist with the following calculations, defined as
\be 
\Z\equiv3(\dot{\psi}+H\phi)-\frac{\nabla^2}{a^2}\sigma\,,
\ee
we can rewrite Eq.~(\ref{eq:ijtrace}), using Eqs.~(\ref{eq:00})
and (\ref{eq:ijtracefree}), as
\be 
\label{eq:zev}
\dot{\Z}+2H\Z+\Big(3\dot{H}+\frac{\nabla^2}{a^2}\Big)\phi
=4\pi G(\delta\rho+3\delta P)
\,.
\ee
Then, from Eq.~(\ref{eq:fluidev}),
\be 
\label{eq:zrel}
\Z=\dot{\dm}+3H\phi+\frac{\nabla^2}{a^2}V_{\rm c}\,,
\ee 
and from Eq.~(\ref{eq:mmtmcons}) for the dark matter fluid (for which $\cs=0$),
\be 
\label{eq:phirel}
\phi=-\dot{V_{\rm c}}\,.
\ee 
Differentiating Eq.~(\ref{eq:zrel}) gives
\be 
\dot{\Z}=\ddot{\dm}+3(H\phi)\dot{}+\frac{\nabla^2}{a^2}
\Big( \dot{V_{\rm c}}-2H V_{\rm c}\Big)\,.
\ee
Substituting this into Eq.~(\ref{eq:zev}) gives 
\begin{align}
\label{eq:fluidmaster}
&\ddot{\dm}+2H\dot{\dm}-4\pi G \rho_{\rm c}\dm
=
8\pi G(2\dot{\vp}\dot{\dvp}-U_{,\vp}\dvp)\nn\\
&\qquad\qquad
+\dot{V_{\rm c}}\Big(6(\dot{H}+H^2)+16\pi G \dot{\vp}^2\Big)+3H\ddot{V_{\rm c}}\,.
\end{align}
The evolution equation for the field is then obtained solely in terms of
matter perturbations from Eq.~(\ref{eq:fieldev})
by using Eqs.~(\ref{eq:phirel}) and (\ref{eq:zrel}):
\bea
\label{eq:fieldmaster}
&&\ddot{\dvp}+3H\dot{\dvp}+\Big(U_{,\vp\vp}-\frac{\nabla^2}{a^2}\Big)\dvp\\
&&\qquad=\dot{\vp}\Big(\dot{\dm}+\frac{\nabla^2}{a^2}V_{\rm c}-\ddot{V_{\rm c}}\Big)
+2U_{,\vp}\dot{V_{\rm c}}\,.\nn
\eea

It is worth restating that we have derived these equations in a general form without fixing a gauge.
If we set the dark matter velocity to zero, i.e. $V_{\rm c}=0$, then they reduce to those
presented in, for example, 
Refs.~\cite{Hwang:2001fb, Chongchitnan:2008ry} (for the case of a zero energy-momentum transfer).

\section{Gauge transformations}
\label{sec:gaugetrans}

As mentioned above, so far the equations have been presented without choosing a gauge. In order to choose a gauge,
 we need to first consider how the variables change under a gauge transformation, that is,
 a change of coordinates in the physical spacetime while holding the background
 coordinates fixed. Recall that, in general
 the gauge transformation of an arbitrary tensor field ${\mathbf T}$ is given by the exponential map
 \cite{MM2008, MW2008}
 \be 
 \widetilde{\mathbf T}=e^{\pounds_{\xi}}{\mathbf T}\,.
 \ee
 This can then be expanded to give the gauge transformation at linear order in perturbation theory
 as
 \be 
 \widetilde{\delta {\mathbf T}}=\delta {\mathbf T}+\pounds_{\xi}{\mathbf T}_0\,,
 \ee 
 where ${\mathbf T}_0$ is the value of the tensor field ${\mathbf T}$ in the background and
 $\pounds_{\xi}$ denotes the Lie derivative with respect to the gauge transformation
  generating vector, $\xi^\mu$. This can be split up as
 \be 
 \xi^\mu=(\alpha, \beta_{,}{}^i)
 \,,
 \ee
 such that the coordinates transform as
 \begin{align}
 \widetilde{t}&=t-\alpha\,,\\
 \widetilde{x}^i&=x^i-\beta_,{}^i\,.
 \end{align}
 Then, one can show that scalar quantities such as the field perturbation transform as
 \be 
 \label{eq:fieldtrans}
 \widetilde{\dvp}=\dvp+\dot{\vp}\alpha\,,
 \ee
 the density contrast transforms as
 \be 
 \widetilde{\dm}=\dm+\frac{\dot{\rho_{\rm c}}}{\rho_{\rm c}}\alpha\,,
 \ee 
and the components of the velocity potential as
\be 
\label{eq:veltrans}
\widetilde{v}_\alpha=v_\alpha-a\dot{\beta}\,.
\ee
Furthermore, by considering the transformation behaviour of the metric
tensor gives us the following transformation rules for the scalar metric
perturbations
\begin{align}
\widetilde{\phi}&=\phi+\dot{\alpha} \,,\\
\widetilde{\psi}&=\psi-H\alpha\,, \\
\widetilde{B}&=B-\frac{1}{a}\alpha+a\dot{\beta}\,,\\
\label{eq:transE}
\widetilde{E}&=E+\beta\,,
\end{align}
and so the scalar shear transforms as
\be 
\widetilde{\sigma}=\sigma+\alpha\,.
\ee
Finally Eq.~(\ref{eq:veltrans}) along with Eq.~(\ref{eq:covvel}), gives the transformation behaviour of
the components of the covariant velocity potential
\be 
\widetilde{V}_\alpha=V_\alpha-\alpha\,.
\ee

\section{discussion and conclusions}
\label{eq:conc}

When splitting the spacetime into a background and a perturbation, as
is done in cosmological perturbation theory, one introduces
a gauge problem. That is, there exist spurious gauge modes which
 must be removed in order to draw physically meaningful conclusions.
 This 
problem arises because, while general relativity is a fully covariant theory, 
this splitting is not a covariant process (see, e.g., Ref.~\cite{MM2008}). In order to resolve this issue Bardeen
proposed a systematic method of dealing with the gauge modes, by
considering the transformation behaviour of quantities and constructing
gauge invariant variables \cite{Bardeen:1980kt}. We follow that approach here,
by considering the gauge transformations of perturbations presented above in 
Section \ref{sec:gaugetrans}
and making gauge choices such that we can remove the components of the 
generating vector $\alpha$ and $\beta$. Any remaining quantities will then 
be gauge invariant.\\

%
%So, now we can consider the main question of the paper: we know that
%the dark energy perturbations are non-negligible, as per the numerical 
%calculations \cite{Hwang:2009zj}, but are there any gauge issues? \\
%
%The first thing to notice is that since Eqs.~(\ref{eq:fluidmaster}) and (\ref{eq:fieldmaster})
%are written in terms of quantities that are spatially gauge invariant, we do not
%need to worry about fixing $\beta$. 
%{\tt [Is this true? Hwang and Noh seem to imply that it is]}\\

Turning now to the case at hand, choosing the gauge in which the perturbation in the 
dark energy field is zero, $\widetilde{\dvp}=0$,
fixes $\alpha$ as 
\be 
\alpha=-\frac{\dvp}{\dot{\vp}}\,.
\ee
Since none of the gauge transformations of the quantities involved in the governing
equations depend upon $\beta$, we do not need to consider fixing this explicitly
here. (Of course, one can rigorously fix $\beta$ by choosing a suitable gauge condition, for example, 
by setting $\widetilde{E}=0$.)
The governing equations in this gauge are then
\begin{align}
\label{eq:un1}
&\ddot{\widehat\dm}+2H\dot{\widehat\dm}-4\pi G \rho_{\rm c}\widehat{\dm}\nn\\
&\qquad\qquad=
\dot{\widehat{V_{\rm c}}}\Big(6(\dot{H}+H^2)+16\pi G \dot{\vp}^2\Big)+3H\ddot{\widehat{{V_{\rm c}}}}\\
\label{eq:un2}
&\ddot{\widehat{V_{\rm c}}}-\frac{2U_{,\vp}}{\dot{\vp}}\dot{\widehat{V_{\rm c}}}
-\frac{\nabla^2}{a^2}\widehat{V_{\rm c}}=\dot{\widehat\dm}\,,
\end{align}
where the hat denotes that the variables are evaluated in the uniform field fluctuation gauge. That is, in this gauge, $\widehat{\dm}$
and $\widehat{V_{\rm c}}$ are gauge invariant variables  defined as
\begin{align}
\widehat{\dm}=\dm-\frac{\dot{\rho_{\rm c}}}{\rho_{\rm c}\dot{\vp}}\dvp\,,\hspace{1cm}
\widehat{V_{\rm c}}=V_{\rm c}+\frac{\dvp}{\dot{\vp}}\,.
\end{align}

Alternatively, choosing a gauge comoving with the dark matter,
in which $\widetilde{V_{\rm c}}=0$ fixes the generating vector as
\be 
\alpha=V_{\rm c}\,,
\ee
and reduces the governing equations to
\begin{align}
\label{eq:com1}
&\ddot{\bar{\dm}}+2H\dot{\bar{\dm}}-4\pi G \rho_{\rm c}\bar{\dm}
=
8\pi G(2\dot{\vp}\dot{\bar{\dvp}}-U_{,\vp}\bar{\dvp})
\,,\\
\label{eq:com2}
&\ddot{\bar{\dvp}}+3H\dot{\bar{\dvp}}+\Big(U_{,\vp\vp}-\frac{\nabla^2}{a^2}\Big)\bar{\dvp}
=\dot{\vp}\dot{\bar{\dm}}\,,
\end{align}
where the bar denotes variables in the comoving gauge and we have
\begin{align}
\bar{\dm}=\dm+\frac{\rho_{\rm c}}{\dot{\rho_{\rm c}}}V_{\rm c}\,,\hspace{1cm}
\bar{\dvp}=\dvp+\dot{\vp}V_{\rm c}\,.
\end{align}

By studying the above systems of equations, it is evident that choosing 
the dark energy field perturbation to be zero is a well defined choice of gauge,
reducing the governing equations to Eqs.~(\ref{eq:un1}) and (\ref{eq:un2}). 
Then, having done so, we are no longer allowed the freedom to make another
choice of gauge. Alternatively, choosing a gauge comoving with the dark
matter uses up the gauge freedom, and so we are not permitted to neglect
the perturbation in the dark energy field. In fact, doing so will result in
erroneous gauge dependent results. It is clearest to see why this is the case
by considering the set of governing equations. By making our choice of gauge we
are left with a set of equations which is gauge invariant: that is, performing a gauge 
transformation will leave the set of equations unchanged. However, by neglecting the
perturbation in the dark energy field after having chosen the gauge comoving with the
dark matter amounts to setting the right hand side of Eq.~(\ref{eq:com1}) to zero. This 
resulting equation will, in general, then no longer be gauge invariant.

Thus, we conclude that
 the dark energy perturbation must be considered in a system
containing a mixture of dark matter and dark energy. Our result is 
consistent with that of Ref.~\cite{Hwang:2009zj}, though we have shown this 
by simply using the formalism of cosmological perturbation theory 
instead of relying on more involved 
numerical calculations.

%Thus, we conclude with the same result as that of Ref.~\cite{Hwang:2009zj}
%namely that, if choosing to work in a gauge comoving with the dark matter,
%one must include perturbations of the dark energy scalar field 
%in order for the system of equations to make sense. Otherwise, if one chooses
%to set the perturbation to the dark energy field to zero then one no longer has
%the freedom to choose another gauge condition, and doing so will make any
%results calculated gauge dependent.

\acknowledgements
The author is grateful to Ian Huston for valuable comments on a previous version
of this manuscript, and to Karim Malik for stimulating discussions. 
AJC is supported by the Science and Technology Facilities Council (STFC).

%%%%%%%%%%%%%%%%%%%%%%%%%%%%%%

%%%%%%%%%%%%%%%%%%%%%%%%%%%%%%%%%%%%%

\begin{thebibliography}{999}
%%%%%%%%%%%%%%%%%%%%%%%%%%%%%%

%\cite{Komatsu:2010fb}
\bibitem{WMAP7}
  E.~Komatsu {\it et al.},
  %``Seven-Year Wilkinson Microwave Anisotropy Probe (WMAP) Observations:
  %Cosmological Interpretation,''
  arXiv:1001.4538 [astro-ph.CO].
  %%CITATION = ARXIV:1001.4538;%%

%\cite{Seljak:2006bg}
\bibitem{Seljak:2006bg}
  U.~Seljak, A.~Slosar and P.~McDonald,
  %``Cosmological parameters from combining the Lyman-alpha forest with CMB,
  %galaxy clustering and SN constraints,''
  JCAP {\bf 0610} (2006) 014
  [arXiv:astro-ph/0604335].
  %%CITATION = JCAPA,0610,014;%%
 
  %\cite{Tegmark:2006az}
\bibitem{Tegmark:2006az}
  M.~Tegmark {\it et al.}  [SDSS Collaboration],
  %``Cosmological Constraints from the SDSS Luminous Red Galaxies,''
  Phys.\ Rev.\  D {\bf 74} (2006) 123507
  [arXiv:astro-ph/0608632].
  %%CITATION = PHRVA,D74,123507;%%
  
  %\cite{Reid:2009xm}
\bibitem{Reid:2009xm}
  B.~A.~Reid {\it et al.},
  %``Cosmological Constraints from the Clustering of the Sloan Digital Sky
  %Survey DR7 Luminous Red Galaxies,''
  Mon.\ Not.\ Roy.\ Astron.\ Soc.\  {\bf 404} (2010) 60
  [arXiv:0907.1659 [astro-ph.CO]].
  %%CITATION = MNRAA,404,60;%%


%\cite{Copeland:2006wr}
\bibitem{Copeland:2006wr}
  E.~J.~Copeland, M.~Sami and S.~Tsujikawa,
  %``Dynamics of dark energy,''
  Int.\ J.\ Mod.\ Phys.\  D {\bf 15} (2006) 1753
  [arXiv:hep-th/0603057].
  %%CITATION = IMPAE,D15,1753;%%
  
\bibitem{K}
 K.~Uddin,
Ph.D. Thesis,  University of London  (2009)



%\cite{Kodama:1985bj}
\bibitem{Kodama:1985bj}
  H.~Kodama and M.~Sasaki,
  %``Cosmological Perturbation Theory,''
  Prog.\ Theor.\ Phys.\ Suppl.\  {\bf 78} (1984) 1.
  %%CITATION = PTPSA,78,1;%%

%\cite{Mukhanov:1990me}
\bibitem{Mukhanov:1990me}
  V.~F.~Mukhanov, H.~A.~Feldman and R.~H.~Brandenberger,
  %``Theory of cosmological perturbations. Part 1. Classical perturbations. Part
  %2. Quantum theory of perturbations. Part 3. Extensions,''
  Phys.\ Rept.\  {\bf 215} (1992) 203.
  %%CITATION = PRPLC,215,203;%%

%\cite{Bertschinger:1993xt}
\bibitem{Bertschinger:1993xt}
  E.~Bertschinger,
  %``Cosmological dynamics: Course 1,''
  arXiv:astro-ph/9503125.
  %%CITATION = ASTRO-PH/9503125;%%
  



%\cite{Riotto:2002yw}
\bibitem{Riotto:2002yw}
  A.~Riotto,
  %``Inflation and the theory of cosmological perturbations,''
  arXiv:hep-ph/0210162.
  %%CITATION = HEP-PH/0210162;%%
  
  \bibitem{durrerbook}
    R.~Durrer,
  {\it The Cosmic Microwave Background,}
 Cambridge, UK: Univ.~Pr.~(2008).

%\cite{Huterer:2006mva}
\bibitem{Huterer:2006mva}
  D.~Huterer and E.~V.~Linder,
  %``Separating dark physics from physical darkness: Minimalist modified
  %gravity vs. dark energy,''
  Phys.\ Rev.\  D {\bf 75} (2007) 023519
  [arXiv:astro-ph/0608681].
  %%CITATION = PHRVA,D75,023519;%%









%\cite{Hwang:2009zj}
\bibitem{Hwang:2009zj}
  C.~G.~Park, J.~c.~Hwang, J.~h.~Lee and H.~Noh,
  %``Roles of dark energy perturbations in the dynamical dark energy models: Can
  %we ignore them?,''
  Phys.\ Rev.\ Lett.\  {\bf 103} (2009) 151303
  [arXiv:0904.4007 [astro-ph.CO]].
  %%CITATION = PRLTA,103,151303;%%
  


%\cite{Hwang:2001fb}
\bibitem{Hwang:2001fb}
  J.~c.~Hwang and H.~Noh,
  %``Cosmological perturbations with multiple fluids and fields,''
  Class.\ Quant.\ Grav.\  {\bf 19} (2002) 527
  [arXiv:astro-ph/0103244].
  %%CITATION = CQGRD,19,527;%%
  
  %\cite{Malik:2008im}
\bibitem{MW2008}
  K.~A.~Malik and D.~Wands,
  %``Cosmological perturbations,''
  Phys.\ Rept.\  {\bf 475} (2009) 1
  [arXiv:0809.4944 [astro-ph]].
  %%CITATION = PRPLC,475,1;%%
  
    %\cite{Malik:2004tf}
\bibitem{Malik:2004tf}
  K.~A.~Malik and D.~Wands,
  %``Adiabatic and entropy perturbations with interacting fluids and fields,''
  JCAP {\bf 0502} (2005) 007
  [arXiv:astro-ph/0411703].
  %%CITATION = JCAPA,0502,007;%%
  
  %\cite{Chongchitnan:2008ry}
\bibitem{Chongchitnan:2008ry}
  S.~Chongchitnan,
  %``Cosmological Perturbations in Models of Coupled Dark Energy,''
  Phys.\ Rev.\  D {\bf 79} (2009) 043522
  [arXiv:0810.5411 [astro-ph]].
  %%CITATION = PHRVA,D79,043522;%%






  \bibitem{MM2008}
  K.~A.~Malik and D.~R.~Matravers,
 %``A Concise Introduction to Perturbation Theory in Cosmology,''
  Class.\ Quant.\ Grav.\  {\bf 25}, 193001 (2008)
  [arXiv:0804.3276 [astro-ph]].
  %%CITATION = CQGRD,25,193001;%%
  







%\cite{Bardeen:1980kt}
\bibitem{Bardeen:1980kt}
  J.~M.~Bardeen,
  %``Gauge Invariant Cosmological Perturbations,''
  Phys.\ Rev.\  D {\bf 22} (1980) 1882.
  %%CITATION = PHRVA,D22,1882;%%


\end{thebibliography}
\end{document}